\documentstyle[11pt,epsf]{article}
 1
 1
 1

\def\lsim{\mathrel{\raise.3ex\hbox{$<$\kern-.75em\lower1ex\hbox{$\sim$}}}}
\def\gsim{\mathrel{\raise.3ex\hbox{$>$\kern-.75em\lower1ex\hbox{$\sim$}}}}
\catcode`\@=11
\def\slash{\mathpalette\make@slash}
\def\make@slash#1#2{\setbox\z@\hbox{$#1#2$}%
  \hbox to 0pt{\hss$#1/$\hss\kern-\wd0}\box0}
\catcode`\@=12 

\begin{document}
\noindent
\thispagestyle{empty}
\renewcommand{\thefootnote}{\fnsymbol{footnote}}
\begin{flushright}
{\bf CERN-TH/99-152}\\
{\bf hep-ph/9905550}\\
{\bf May 1999}\\
\end{flushright}
\vspace{.5cm}
\begin{center}
  \begin{Large}\bf
$1S$ and  $\overline{\mbox{MS}}$ Bottom Quark Masses 
from $\Upsilon$ Sum Rules
  \end{Large}
  \vspace{1.5cm}

\begin{large}
 A.H. Hoang
\end{large}
  \vspace{.2cm}

\begin{center}
\begin{it}
   Theory Division, CERN\\
   CH-1211 Geneva 23, Switzerland\\
\end{it} 
\end{center}

  \vspace{4cm}
  {\bf Abstract}\\
\vspace{0.3cm}
%
\noindent
\begin{minipage}{15.0cm}
\begin{small}
The bottom quark $1S$ mass, $M_b^{1S}$, is determined using sum rules
which relate the masses and the electronic decay widths of the
$\Upsilon$ mesons to moments of the vacuum polarization
function. The $1S$ mass is defined as half the
perturbative mass of a fictitious ${}^3\!S_1$ bottom-antibottom quark
bound state, and is free of the ambiguity of order $\Lambda_{QCD}$
which plagues the pole mass definition. Compared to an earlier
analysis by the same author, which had been carried out in the pole
mass scheme, the $1S$ mass scheme leads to a much better behaved
perturbative series of the moments, smaller uncertainties in the mass
extraction and to a reduced correlation of the mass and the strong
coupling. We arrive at $M_b^{1S}=4.71\pm 0.03$~GeV taking
$\alpha_s(M_Z)=0.118\pm 0.004$ as an input. From that we determine the
$\overline{\mbox{MS}}$ mass as $\overline m_b(\overline m_b) = 4.20
\pm 0.06$~GeV. The error in $\overline m_b(\overline m_b)$ can be
reduced if the three-loop corrections to the relation of pole and
$\overline{\mbox{MS}}$ mass are known and if the error in the strong
coupling is decreased. 
\\[3mm]
PACS numbers: 14.65.Fy, 13.20.Gd, 13.20.Gv.
\end{small}
\end{minipage}
\end{center}
\setcounter{footnote}{0}
\renewcommand{\thefootnote}{\arabic{footnote}}
\vspace{1.2cm}
%
%
%
\newpage
\noindent
\section{Introduction}
\label{sectionintroduction}
The determination of the Cabibbo-Kobayashi-Maskawa (CKM) matrix
elements is one of the main goals of future $B$ physics experiments. A
sufficiently accurate determination of the size of the CKM matrix
elements and their relative phases will lead to a better understanding
of the origin of CP violation, the structure of the weak interaction,
and, possibly, to the establishment of physics beyond the Standard
Model. For the extraction of the CKM matrix elements from inclusive $B$
decay rates, particularly of $V_{cb}$ from the semileptonic decays
into $D$ mesons, an accurate and precise knowledge of the bottom quark
mass is desirable due to the strong dependence of the total decay
rate on the bottom quark mass parameter.

Non-relativistic sum rules for the masses and electronic decay widths
of $\Upsilon$ mesons, bottom-antibottom quark bound states having
photonic quantum numbers, are an ideal tool to determine the bottom
quark mass: using causality and global duality arguments one can
relate integrals over the total cross section for the production of
hadrons containing a bottom-antibottom quark pair in $e^+e^-$
collisions to derivatives of the vacuum polarization function of
bottom quark currents at zero-momentum transfer. For a particular
range of numbers of derivatives the moments are saturated by the
experimental data on the $\Upsilon$ mesons and, at the same time, can
be calculated reliably using perturbative QCD in the non-relativistic
expansion. Because the moments have, for dimensional reasons, a strong
dependence on the bottom quark mass, these sum rules can be used to
determine the bottom quark mass to high precision. Within the last few
years there have been several analyses at NLO and NNLO in the
non-relativistic expansion, where it has been attempted to extract the
bottom quark pole mass from these sum
rules.~\cite{Voloshin1,Kuhn1,Penin1,Hoang1,Melnikov1}\footnote{
In Ref.~\cite{Jamin1} a sum rule analysis has been performed which is
not compatible with the non-relativistic velocity counting rules.
} 
The results for the pole mass given in these analyses vary rather
strongly in their central values and their precision. This reflects
the fact that the concept of the pole mass is
ambiguous to an amount of order the typical hadronization scale
$\Lambda_{QCD}$~\cite{Beneke1,Bigi1}. From a technical point of view
the large
variations in the extracted pole mass values arise from large
correlations to the value of the strong coupling $\alpha_s$, the 
theoretical uncertainties coming from large NNLO corrections to the
moments, a strong residual dependence on the renormalization
scale, and a different interpretation and treatment of these sources
of uncertainties in the various analyses. From a conceptual point of
view at least part of the former issues can be traced back to the fact
that in the pole mass 
scheme the theoretical expressions for the moments are sensitive to
scales which are below the scales characteristic to the
non-relativistic bottom-antibottom quark dynamics encoded in the
moments. This sensitivity to low scales, which becomes stronger at
higher orders of perturbation theory and which is called the ``renormalon''
problem of the pole mass definition, is an artifact of the pole mass
scheme, and does not exist if a mass definition is employed which is
not ambiguous to an amount of order 
$\Lambda_{QCD}$~\cite{Hoang2,Beneke2}. Such
mass definitions are referred to as ``short-distance''
masses. (Short-distance masses can still have ambiguities of order
$\Lambda_{QCD}^2$ divided by the heavy quark mass.) In
general, these mass definitions have a nicely behaved perturbative
relation to the $\overline{\mbox{MS}}$ mass. This means that the
perturbative relation between the short-distance masses and the
$\overline{\mbox{MS}}$ mass can be expected to be convergent for
all practical purposes (i.e. as long as we only deal with a few low
orders of perturbation theory).\footnote{
The words like ``convergent'' and ``nicely behaved'', which are used in
this work to describe the perturbative relations between
short-distance masses, do not have a strict mathematical
meaning. They are merely chosen to distinguish from the situation one
finds if the pole mass is related to a short-distance mass. We would
describe the latter situation with the words ``not convergent'' and
``badly behaved''. From the mathematical point of view, of course, all
series shown in this work are asymptotic. 
} 
However, not any short-distance mass is
equally well adapted to improve the situation for the $\Upsilon$ sum
rules because the correlations to the strong coupling and the strong
dependence on the renormalization scale observed in the pole mass
scheme already exist in the leading order theoretical expressions for
the moments and are therefore not related to the problem of an
increased infrared
sensitivity. Thus, it is advantageous to use a specialised
short-distance mass for the mass extraction from the $\Upsilon$ sum
rules, which eliminates, as much as possible, correlations and
dependences to other parameters and which stabilises the perturbative
expansion of the moments. Such a specialised short-distance mass can
be either used as a mass definition in its own right, or, in a second
step, related to other short-distance masses like the
$\overline{\mbox{MS}}$ mass, which can be regarded as a specialised
short-distance mass designed for high energy processes. In this step
a part of the correlations to the strong coupling are expected to come
back. However, correlations to other parameters like renormalization
scales are eliminated, which can lead to a reduction of uncertainties.
In addition, the perturbative relations between short-distance
masses is well convergent and possible sources of correlations which
arise in these relations can be easily identified. 

In this paper we extract the $1S$ bottom quark mass from the $\Upsilon$
sum rules based on the theoretical expressions of the moments which we
have calculated analytically at next-to-next-to-leading order (NNLO)
in the pole mass scheme in a previous publication~\cite{Hoang1}. In
the $1S$
scheme the bottom quark mass is defined as half the perturbative mass
of a colour singlet bottom-antibottom quark $J^{PC} = 1^{--}$,
${}^3\!S_1$ bound state. The $1S$ mass is a short-distance mass and
has been used previously to parameterise inclusive $B$ mesons decays
leading to a considerable reduction of the 
perturbative corrections.~\cite{Hoang3,Hoang4} 
In the $1S$ mass scheme we find a reduction in the size of the large NNLO
corrections to the theoretical moments observed in the pole mass
scheme, and a much weaker dependence of the moments on $\alpha_s$ and
the renormalization scale governing the non-relativistic
bottom-antibottom quark dynamics. This results in a weak
correlation of the $1S$ mass to the value of the
strong coupling and to much smaller uncertainties in the mass
determination compared to our earlier analysis in the pole scheme
($M_b^{1S}=4.71\pm 0.03$~GeV in this work versus $M_b^{pole}=4.9\pm
0.1$~GeV in Ref.~\cite{Hoang1} if $\alpha_s(M_Z)=0.118\pm 0.004$ is taken
as input) using exactly the same statistical analysis and the same way
to treat theoretical uncertainties. 

By construction, the $1S$ mass is half the perturbative contribution of
the $\Upsilon(1S)$ meson mass ($M_{\Upsilon(1S)}=9460.37\pm
0.21$~MeV~\cite{PDG}). In Refs.~\cite{Hoang3,Hoang4} this relation has
been used
explicitly, and an uncertainty of $\pm 50$~MeV has been assigned to the
value of $M_b^{1S}$ based on conservative general arguments to estimate
the size of non-perturbative effects in the $\Upsilon(1S)$ mass. In
this work this relation is not used and
$M_b^{1S}$ is treated as a fictitious mass parameter which
is determined solely from the sum rule analysis. The close proximity
of $M_b^{1S}$ determined in this work and
$M_{\Upsilon(1S)}/2=4.730$~GeV is 
non-trivial because higher radial $\Upsilon$ meson excitations and the
$b\bar b$ continuum have a significant contribution to the sum rule.
Thus the analysis in this work provides an independent quantitative
cross check of the arguments about the size of non-perturbative
effects in the $\Upsilon(1S)$ mass made in Refs.~\cite{Hoang3,Hoang4}.

The program of this paper is as follows: in Sec.~\ref{section1Smass} 
we introduce the $1S$ mass. In Sec.~\ref{sectionsumrule} we briefly
review the $\Upsilon$ sum rules and discuss how the analytic results
obtained in Ref.~\cite{Hoang1} in the pole scheme are modified in the
$1S$ mass scheme. In Sec.~\ref{sectionnumerics} the statistical
analysis is
explained and the results of the determination of the $1S$ bottom
quark mass are shown. In Sec.~\ref{sectionmsbar} we discuss the
upsilon expansion~\cite{Hoang3,Hoang4} which is needed to relate the
$1S$ mass to other short-distance mass definitions, and we determine
the $\overline{\mbox{MS}}$ mass. In Sec.~\ref{sectionother} we determine
the value of other recently proposed specialised short-distance masses
which can be used in the $\Upsilon$ sum rules, and
Sec.~\ref{sectionconclusions} contains the conclusions.

\par
\vspace{0.5cm}
\section{The $1S$ Bottom Quark Mass}
\label{section1Smass}
The $1S$ bottom quark mass is defined as half the
perturbative mass of a  $J^{PC} = 1^{--}$,
${}^3\!S_1$ bottomonium ground state. Expressed in terms of the
pole mass the $1S$ mass at NNLO in the non-relativistic
expansion reads ($a_s=\alpha_s(\mu)$),~\cite{Pineda1,Melnikov1} 
\begin{eqnarray}
M_b^{1S} & = & M_b^{pole}\,\bigg[\,1 \,-\,
\epsilon\,\Delta^{\mbox{\tiny LO}}(a_s)
\,-\,\epsilon^2\,\Delta^{\mbox{\tiny NLO}}(M_b^{pole},a_s,\mu)
\,-\,\epsilon^3\,\Delta^{\mbox{\tiny NNLO}}(M_b^{pole},a_s,\mu)
\,\bigg]
\,,
\label{M1Sdef}
\end{eqnarray}
where
\begin{eqnarray}
\Delta^{\mbox{\tiny LO}} & = &
 \frac{C_F^2\,a_s^2}{8}
\,,
\label{deltaLO}
\\[2mm]  
\Delta^{\mbox{\tiny NLO}} & = &
\frac{C_F^2\,a_s^2}{8}\, 
\Big(\frac{a_s}{\pi}\Big)\,\bigg[\,
\beta_0\,\bigg( L + 1 \,\bigg) + \frac{a_1}{2} 
\,\bigg]
\,,
\label{deltaNLO}
\\[2mm] 
\Delta^{\mbox{\tiny NNLO}} & = &
\frac{C_F^2\,a_s^2}{8}\, \Big(\frac{a_s}{\pi}\Big)^2\,
\bigg[\,
\beta_0^2\,\bigg(\, \frac{3}{4} L^2 +  L + 
                             \frac{\zeta_3}{2} + \frac{\pi^2}{24} +
                             \frac{1}{4} 
\,\bigg) + 
\beta_0\,\frac{a_1}{2}\,\bigg(\, \frac{3}{2}\,L + 1
\,\bigg)
\nonumber\\[3mm]
& & \hspace{1.5cm} +
\frac{\beta_1}{4}\,\bigg(\, L + 1
\,\bigg) +
\frac{a_1^2}{16} + \frac{a_2}{8} + 
\bigg(\, C_A - \frac{C_F}{48} \,\bigg)\, C_F \pi^2 
\,\bigg]
\,,
\label{deltaNNLO}
\\[2mm]
L & \equiv & \ln\Big(\frac{\mu}{C_F\,a_s\,M_b^{pole}}\Big)
\,,
\end{eqnarray}
and ($n_l=4$)
\begin{eqnarray}
\beta_0 & = & \frac{11}{3}\,C_A - \frac{4}{3}\,T\,n_l
\,,
\nonumber\\[2mm]
\beta_1 & = & \frac{34}{3}\,C_A^2 
-\frac{20}{3}C_A\,T\,n_l
- 4\,C_F\,T\,n_l
\,,
\nonumber\\[2mm]
a_1 & = &  \frac{31}{9}\,C_A - \frac{20}{9}\,T\,n_l
\,,
\nonumber\\[2mm]
a_2 & = & 
\bigg(\,\frac{4343}{162}+4\,\pi^2-\frac{\pi^4}{4}
 +\frac{22}{3}\,\zeta_3\,\bigg)\,C_A^2 
-\bigg(\,\frac{1798}{81}+\frac{56}{3}\,\zeta_3\,\bigg)\,C_A\,T\,n_l
\nonumber\\[2mm] & &
-\bigg(\,\frac{55}{3}-16\,\zeta_3\,\bigg)\,C_F\,T\,n_l 
+\bigg(\,\frac{20}{9}\,T\,n_l\,\bigg)^2
\,.
\end{eqnarray}
The constants $\beta_0$ and $\beta_1$ are the one- and two-loop
coefficients of the QCD beta function and the constants
$a_1$~\cite{Fischler1,Billoire1} and $a_2$~\cite{Schroeder1,Peter1}
the non-logarithmic one- and two-loop corrections to the static
colour-singlet heavy quark potential in the pole mass scheme,
$V^{Coul}$.\footnote{ 
The constant $a_2$ was first calculated in Ref.~\cite{Peter1}.
Recently an error in the coefficient of the term
$\propto \pi^2 C_A^2$ was corrected in
Ref.~\cite{Schroeder1}. Although this leads to a
reduction of the value of $a_2$ by a factor of two, the change turns
out to be irrelevant for the mass determination in this work, because
the influence of $a_2$ is, by construction, strongly suppressed in the
$1S$ scheme.
} 
The charm quark is treated as massless.
In Eq.~(\ref{M1Sdef}) we have labelled the contributions at LO, NLO and
NNLO in the non-relativistic expansion by powers $\epsilon$,
$\epsilon^2$ and $\epsilon^3$, respectively, of the auxiliary
parameter $\epsilon=1$. The expansion in terms of the parameter
$\epsilon$ is called the upsilon expansion~\cite{Hoang3,Hoang4} and
will be 
relevant when the $1S$ is related to the $\overline{\mbox{MS}}$
mass. We will come back to this issue in Sec.~\ref{sectionmsbar}

In the framework of the non-relativistic power counting scheme
$\Delta_{LO}$, $\Delta_{NLO}$ and $\Delta_{NNLO}$ are of order $v^2,
v^3$ and $v^4$ respectively.
In order to implement the $1S$ mass into the analytic expressions for
the moments in the pole mass scheme we have to invert
relation~(\ref{M1Sdef}) in the non-relativistic framework,
\begin{eqnarray}
M_b^{pole} & = &
M_b^{1S}\,\bigg\{\,
1 \,+\,
\Delta^{LO}(a_s)
\,+\,\Delta^{NLO}(M_b^{1S},a_s,\mu)
\nonumber
\\[3mm] & &
\mbox{\hspace{1cm}} 
\,+\,
\bigg[\,\Delta^{NNLO}(M_b^{1S},a_s,\mu)
+\Big(\Delta^{LO}(a_s)\Big)^2\,\bigg]
\,\bigg\}
\,.
\label{MpoleM1S}
\end{eqnarray} 
The terms in the brackets on the RHS of Eq.~(\ref{MpoleM1S}) represent
the NNLO contributions.
We would like to emphasise again that $M_b^{1S}$ is a short-distance
mass because it does not suffer from the ambiguity of order
$\Lambda_{QCD}$ like the pole mass. This is because the $1S$ mass
contains, by construction, the half of the total static energy 
$\langle 2 M_b^{pole} + V^{Coul}\rangle$ which
can be proven to be ambiguity-free at order
$\Lambda_{QCD}$.~\cite{Hoang2,Beneke2} 
\par
\vspace{0.5cm}
\section{The $\Upsilon$ Sum Rules}
\label{sectionsumrule}
In this section we briefly review the basic concepts involved in the
$\Upsilon$ meson sum rules. We outline the calculation of the moments
in the pole mass scheme which has been carried out in
Ref.~\cite{Hoang1}, and we describe how the moments have to be
modified in the $1S$ scheme. We will not present any details on the
original computations carried out in the pole mass scheme and refer
the interested reader to Ref.~\cite{Hoang1}.

The sum rules for the $\Upsilon$ mesons start from the correlator of
two electromagnetic currents of bottom quarks at momentum transfer
$q$,
\begin{equation}
\Pi_{\mu\nu}(q) \, = \,
-\,i \int \!{\rm d}x\,e^{i\,q.x}\,
   \langle\, 0\,|\,T\,j^b_\mu(x)\,j^b_\nu(0)\,|\,0\, \rangle
\,,
\label{vacpoldef}
\end{equation}
where
\begin{equation}
j^b_\mu(x) \, =  \bar b(x)\,\gamma_\mu\,b(x)
\,,
\end{equation}
and the symbol $b$ denotes the bottom quark Dirac field. The $n$-th
moment $P_n$ of the vacuum polarization function is defined as
\begin{equation}
P_n \, \equiv \,
\frac{4\,\pi^2\,Q_b^2}{n!\,q^2}\,
\bigg(\frac{d}{d q^2}\bigg)^n\,\Pi_\mu^{\,\,\,\mu}(q)\bigg|_{q^2=0}
\,,
\label{momentsdef1}
\end{equation}
where $Q_b=-1/3$ is the electric charge of the bottom quark.
Due to causality the n-th moment $P_n$ can also be written as a
dispersion integral
\begin{equation}
P_n \, = \,
\int\limits^\infty_{\sqrt{s}_{min}} \frac{d s}{s^{n+1}}\,R(s)
\,,
\label{momentsdef2}
\end{equation}
where
\begin{equation}
R(s) \, = \, \frac{\sigma(e^+e^-\to\gamma^*\to \mbox{``$b\bar b$''})}
{\sigma_{pt}}
\label{Rdefinitioncovariant}
\end{equation}
is the total photon mediated cross section of bottom quark-antiquark
production in $e^+e^-$ annihilation normalised to the point cross
section $\sigma_{pt}\equiv 4 \pi \alpha^2/3 s$, and $s$ the square of the
centre-of-mass energy. The lower limit of the
integration in Eq.~(\ref{momentsdef2}) is set by the mass of the
lowest lying resonance. Assuming global duality the moments $P_n$ can
be either calculated from experimental data on $R$ or theoretically
using perturbative QCD. 

The experimental moments $P_n^{ex}$ are determined using latest data
on the $\Upsilon$ meson masses, $M_{\Upsilon(nS)}$, and electronic
decay widths, $\Gamma_{\Upsilon(nS)}$, for $n=1,\ldots,6$. The formula
for the experimental moments used in this work reads
\begin{eqnarray}
P_n^{ex} & = &  
\frac{9\,\pi}{\tilde\alpha^2_{em}}\,\sum\limits_{k=1}^6\,
\frac{\Gamma_{\Upsilon(kS)}}{M_{\Upsilon(k)}^{2n+1}} 
\, + \,
\int\limits_{\sqrt{s}_{B\bar B}}^\infty
\frac{ds}{s^{n+1}}\,r_{cont}(s)
\,.
\label{Pnexperiment}
\end{eqnarray}
and is based on the narrow width approximation for the known
$\Upsilon$ resonances. $\tilde\alpha_{em}$ is the electromagnetic
coupling at the scale $10$~GeV. Because the difference in the
electromagnetic coupling for the different $\Upsilon$ masses is
negligible we chose $10$~GeV as the scale of the electromagnetic
coupling for all resonances.
The continuum cross section above the $B\bar B$
threshold is approximated by the constant $r_c=1/3$, which is equal to
the born cross section for $s\to\infty$, assuming a $50\%$
uncertainty\footnote{
We take the opportunity to point out a typo in Eq.~(77) of
Ref.~\cite{Hoang1}, where a factor $1/3$ is missing. This typo only
exists in the text of Ref.~\cite{Hoang1} and is not contained in the
numerical codes. 
} 
\begin{equation}
r_{cont}(s) = r_c\,(1 \pm 0.5)
\,.
\label{Rexperimentcontinuum}
\end{equation}
For $n\ge 4$ the continuum contribution is already sufficiently
suppressed that a more detailed description is not needed.
For a compilation of all experimental numbers used in this work see
Tab.~\ref{tabdata}.

\begin{table}[t!]  
\vskip 7mm
\begin{center}
\begin{tabular}{|l||c|c|c|} \hline
\multicolumn{1}{|c||}{$nS$} 
  & $1S$ & $2S$ & $3S$ 
 \\ \hline\hline 
$M_{nS}/[\mbox{GeV}]$  
 & $9.460$ & $10.023$ & $10.355$    \\ \hline
$\Gamma_{nS}/[\mbox{keV}]$ 
 & $1.32\pm 0.04\pm 0.03$ & $0.52\pm 0.03\pm 0.01$ 
 & $0.48\pm 0.03\pm 0.03$ \\ \hline\hline\hline
\multicolumn{1}{|c||}{$nS$} 
  & $4S$ & $5S$ & $6S$ 
 \\ \hline\hline 
$M_{nS}/[\mbox{GeV}]$  
 & $10.58$ & $10.87$ & $11.02$   \\ \hline
$\Gamma_{nS}/[\mbox{keV}]$ 
 & $0.25\pm 0.03\pm 0.01$ 
 & $0.31\pm 0.05\pm 0.07$ & $0.13\pm 0.03\pm$ 0.03  \\ \hline\hline\hline
\multicolumn{4}{|c|}
{$\tilde \alpha_{em}^{-1} = \alpha_{em}^{-1}(10$~GeV$) = 131.8(1\pm0.005)\,,
\quad (\sqrt{s})_{B\bar B} = 2\times 5.279$~GeV} \\ \hline
\end{tabular}
\caption{\label{tabdata} 
The experimental numbers for the $\Upsilon$ masses and electronic decay
widths used for the calculation of the experimental moments
$P_n^{ex}$. For the widths the first error is statistical and the
second systematical. The errors for the partial widths of
$\Upsilon_{1S}$ and $\Upsilon_{2S}$ are taken from
Ref.~\cite{Albrecht1}. All the other errors are estimated from the
numbers presented in Ref.~\cite{PDG}. The small errors in the
$\Upsilon$ masses and the $B\bar B$ threshold $(\sqrt{s})_{B\bar B}$
are neglected.
}
\end{center}
\vskip 3mm
\end{table}

A reliable computation of the theoretically moments based on
perturbative QCD is only possible if the effective energy range
contributing to the integration in Eq.~(\ref{momentsdef2}) is
sufficiently larger than
$\Lambda_{QCD}\sim{\cal{O}}(200-300~\mbox{MeV})$~\cite{Poggio1}. For
large values of $n$ one can show that the size of this
energy range is of order $M_b/n$. This means 
that $n$ should be chosen sufficiently smaller than $15-20$. To
suppress systematic theoretical uncertainties as much as possible we
take $n_{max}=10$ as the maximal allowed value for $n$. However, it is
also desirable to choose $n$ as large as possible in order to suppress
the contribution from the $b\bar b$ continuum to
$R(s)$ above the $B\bar B$ threshold, which is rather poorly known
experimentally.
In other words, one has to choose $n$ large enough that the
bottom-antibottom quark dynamics encoded in the moments $P_n$ is
non-relativistic. Because the effective size of the energy range
contributing to the $n$-th moment is of order $M_b/n$, the mean
centre-of-mass velocity of the bottom quarks in the $n$-th moment is
of order
\begin{equation}
v \, \sim \, \alpha_s \, \sim \, \frac{1}{\sqrt{n}}
\,,
\label{powercounting}
\end{equation} 
where $v\sim\alpha_s$ is characteristic for perturbative
non-relativistic quark-antiquark systems.
In our analysis we choose $n_{min}=4$ as the minimal value of $n$ to
ensure the dominance of the non-relativistic dynamics in the moments.
For the values of $n$ employed in our analysis the non-perturbative
contributions coming from the gluon condensate are at the per-mill
level and negligible.~\cite{Voloshin1} 

In Ref.~\cite{Hoang1} we have used non-relativistic QCD (NRQCD) as
formulated in Ref.~\cite{Caswell1} to determine the theoretical moments
$P_n^{th}$ in the pole mass scheme at NNLO in the non-relativistic
expansion, which included all corrections up to order
$1/n$, $\alpha_s/\sqrt{n}$ and $\alpha_s^2$ with respect to the
expressions which are leading order in the non-relativistic expansion.
Taking into account the power counting shown in
Eq.~(\ref{powercounting}) the dispersion integration for the
theoretical moments $P_n^{th}$ at NNLO takes the form
\begin{equation}
P_n^{th} \, = \,
\frac{1}{4^{n}\,(M_b^{pole})^{2n}}\,\int\limits_{E_{\rm bind}}^\infty 
\frac{d E}{M_b^{pole}} \,\exp\bigg\{\,
-\frac{E}{M_b^{pole}}\,n
\,\bigg\}\,\bigg(\,
1 - \frac{E}{2\,M_b^{pole}} + \frac{E^2}{4\,(M_b^{pole})^2}\,n
\,\bigg)\,R_{\mbox{\tiny NNLO}}^{\mbox{\tiny thr}}(E)
\,,
\label{Pnexpression1}
\end{equation}
where $E\equiv\sqrt{s}-2M_b^{pole}$ and $E_{\rm bind}$ is the binding
energy of the lowest lying resonance, i.e. 
$E_{\rm bind}=2(M_b^{1S}-M_b^{pole})$. The exponential form of the LO
non-relativistic contribution to the energy integration has to be
chosen because $E$ scales like $v^2\sim 1/n$.
The result for the theoretical moments at NNLO can be cast into the
form
\begin{eqnarray}
P_n^{th} & = &
\frac{3\,N_c\,Q_b^2\,\sqrt{\pi}}{4^{n+1}\,(M_b^{pole})^{2n} \, n^{3/2}}\,
\bigg\{\,
C_1\Big(\frac{\mu_{\rm hard}}{M_b^{pole}},\frac{\mu_{\rm
    fac}}{M_b^{pole}},\alpha_s(\mu_{\rm hard})\Big)\,
\varrho_{n,1}\Big(\frac{\mu_{\rm soft}}{M_b^{pole}},
    \frac{\mu_{\rm fac}}{M_b^{pole}},\alpha_s(\mu_{\rm soft})\Big)
\nonumber\\[2mm] & &
\mbox{\hspace{4cm}}
 +\, \varrho_{n,2}\Big(\alpha_s(\mu_{\rm soft})\Big)
\,\bigg\}
\,,
\label{Pntheoryfinal}
\end{eqnarray}
where $\varrho_{n,1}$ describes the contribution to the moments coming
from the dominant non-relativistic current correlator involving two
dimension three ${}^3\!S_1$ NRQCD currents, and $\varrho_{n,2}$ the
contribution coming from the NNLO current correlator involving one
dimension three and one dimension five ${}^3\!S_1$ NRQCD current. $C_1$
contains the short-distance corrections to the dimension three
currents up to order ${\cal{O}}(\alpha_s^2)$. The corresponding
short-distance correction to $\varrho_{n,2}$ is not needed because the
latter is already of NNLO. In Eq.~(\ref{Pntheoryfinal}) we have
indicated the dependence of the moments on the various
renormalization scales used in Ref.~\cite{Hoang1}. $\mu_{\rm soft}$
is the renormalization scale 
of the strong coupling governing the non-relativistic dynamics and
$\mu_{\rm hard}$ the scale of the strong coupling in the
short-distance coefficient $C_1$. The scale $\mu_{\rm fac}$ is the
factorisation scale which separates non-relativistic and
short-distance momenta. The three scales arise in the calculation of
Ref.~\cite{Hoang1} as a consequence of the use of a cutoff-like
regularization for the UV divergences which arise in NRQCD Feynman
diagrams, while the strong coupling is renormalised in the
$\overline{\mbox{MS}}$ scheme. Formally the moments are invariant
under changes of these scales at NNLO. From the numerical point of
view, however, they are not, because the dependence on the scales, in
particular the soft scale, only cancels partially at finite order of
perturbation theory. In the statistical analysis for the bottom mass
determination all three scales are varied independently in order to
estimate theoretical uncertainties. Eq.~(\ref{Pntheoryfinal}) also
displays the dependence on the pole mass. The
short-distance factor $C_1$ and $\varrho_{n,1}$ only depend on the
pole mass through the logarithm of the ratios of the renormalization
scales and the pole mass, which originate from the running of the strong
coupling and the NRQCD UV divergences. The latter dependences only
arise at NLO and NNLO and do not lead to any modifications in the $1S$
scheme because the difference between $1S$ and pole mass is of order
$v^2\sim\alpha_s^2\sim 1/n$ in the non-relativistic expansion. The
most important pole mass dependence is the overall factor
$(M_b^{pole})^{-2n}$. In the $1S$ scheme this factor reads 
\begin{eqnarray}
\lefteqn{
\frac{1}{(M_b^{pole})^{2n}} \, = \,
\frac{1}{(M_b^{1S})^{2n}}\,
\exp\Big(-2\,n\,\Delta^{\mbox{\tiny LO}}(\alpha_s(\mu_{\rm soft}))\Big)\,
\bigg\{\,
1 - 2\,n\,\Delta^{\mbox{\tiny NLO}}(M_b^{1S},\alpha_s(\mu_{\rm soft}),\mu_{\rm
  soft}) 
}
\nonumber\\[2mm] & &
\mbox{\hspace{0cm}}
+\,n\bigg[
\Big(\Delta^{\mbox{\tiny LO}}(\alpha_s(\mu_{\rm soft}))\Big)^2 - 
2\Delta^{\mbox{\tiny NNLO}}(M_b^{1S},\alpha_s(\mu_{\rm soft}),\mu_{\rm soft}) +
2n\Big(\Delta^{\mbox{\tiny NLO}}(M_b^{1S},\alpha_s(\mu_{\rm soft}),\mu_{\rm
  soft})\Big)^2
\bigg]
\bigg\}
\nonumber\\
\label{finalfactor}
\end{eqnarray}
at NNLO in the non-relativistic expansion using the relation given in
Eq.~(\ref{MpoleM1S}).
\begin{table}[t] 
\vskip 7mm
\begin{center}
\begin{tabular}{|r@{$/$}l||c|c|c||c|c|c|} \hline
\multicolumn{2}{|c||}{Moment} 
   & \multicolumn{3}{|c||}{$M_b^{1S} (M_b^{pole})/[GeV]$}
   & \multicolumn{3}{|c|}{$\alpha_s(M_Z)$} \\ \hline
\multicolumn{2}{|c||}{}
 & $4.6$ &  $4.8$ & $5.0$ & $0.11$ & $0.12$
                          & $0.13$ \\ \hline\hline 
$P_4^{th}$&$[10^{-8}\,\mbox{GeV}^{-8}]$
 & $0.27$($0.39$) & $0.19$($0.28$) & $0.14$($0.20$)  
 & $0.17$($0.22$) & $0.20$($0.30$) & $0.26$($0.46$)   \\ \hline
$P_6^{th}$&$[10^{-12}\,\mbox{GeV}^{-12}]$
 & $0.28$($0.47$) & $0.17$($0.29$) & $0.11$($0.18$) 
 & $0.14$($0.21$) & $0.18$($0.31$) & $0.24$($0.54$)  \\ \hline
$P_8^{th}$&$[10^{-16}\,\mbox{GeV}^{-16}]$
 & $0.33$($0.64$) & $0.17$($0.33$) & $0.09$($0.17$)
 & $0.14$($0.22$) & $0.18$($0.36$) & $0.25$($0.70$)  \\ \hline
$P_{10}^{th}$&$[10^{-20}\,\mbox{GeV}^{-20}]$
 & $0.41$($0.90$) & $0.18$($0.39$) & $0.08$($0.17$)
 & $0.14$($0.25$) & $0.19$($0.44$) & $0.27$($0.94$)  \\ \hline
$P_{20}^{th}$&$[10^{-40}\,\mbox{GeV}^{-40}]$
 & $1.64$($6.85$) & $0.30$($1.26$) & $0.06$($0.25$) 
 & $0.22$($0.63$) & $0.33$($1.52$) & $0.53$($4.72$) \\ \hline\hline
\multicolumn{2}{|c||}{} 
   & \multicolumn{3}{c||}{\raisebox{-2.5ex}[2.5ex]{$\alpha_s(M_Z) = 0.118$}}
   & \multicolumn{3}{c|}{\raisebox{-.5ex}[.5ex]{$M_b^{1S}=4.8$~GeV}}\\
\multicolumn{2}{|c||}{} 
   & \multicolumn{3}{c||}{}
   &
   \multicolumn{3}{c|}{\raisebox{.5ex}[-.5ex]{($M_b^{pole}=4.8$~GeV)}} \\ \hline
\multicolumn{2}{|c||}{} 
   &      \multicolumn{6}{c|}{$\mu_{\rm soft}=2.5\,\mbox{GeV}\,,\quad
          \mu_{\rm hard}=\mu_{\rm fac}=5$~GeV} \\\hline
\end{tabular}
\caption{\label{tabcomp1} 
The theoretical moments $P_n^{th}$ at NNLO in the $1S$ (pole) mass
scheme for $n=4,6,8,10,20$ and fixed $\mu_{\rm soft}=2.5$~GeV and 
$\mu_{\rm hard}=\mu_{\rm fac}=5$~GeV for various values of $M_b^{1S}$
($M_b^{pole}$) and $\alpha_s(M_z)$. The two-loop running  has been
employed for the strong coupling. The values for $P_n^{th}$ in the pole
mass scheme are slightly different from the numbers shown in
Ref.~\cite{Hoang1} because of the correction in  $a_2$.
 }
\end{center}
\vskip 3mm
\end{table}
\begin{table}[htb] 
\vskip 7mm
\begin{center}
\begin{tabular}{|r@{$/$}l||c|c||c|c||c|c|} \hline
\multicolumn{2}{|c|}{Moment} 
   & \multicolumn{2}{c||}{$\mu_{\rm soft}/[\mbox{GeV}]$}
   & \multicolumn{2}{c||}{$\mu_{\rm hard}/[\mbox{GeV}]$}
   & \multicolumn{2}{c|}{$\mu_{\rm fac}/[\mbox{GeV}]$} \\ \hline
\multicolumn{2}{|c|}{}
 & $1.5$ & $3.5$ & $2.5$ & $10.0$ 
                          & $2.5$ & $10.0$ \\ \hline\hline
$P_4^{th}$&$[10^{-8}\,\mbox{GeV}^{-8}]$
 & $0.33$($0.51$) & $0.17$($0.23$)
 & $0.17$($0.25$) & $0.21$($0.30$)
 & $0.21$($0.31$) & $0.16$($0.22$) \\ \hline
$P_6^{th}$&$[10^{-12}\,\mbox{GeV}^{-12}]$
 & $0.31$($0.57$) & $0.14$($0.22$)
 & $0.15$($0.26$) & $0.19$($0.31$)
 & $0.19$($0.33$) & $0.14$($0.22$) \\ \hline
$P_8^{th}$&$[10^{-16}\,\mbox{GeV}^{-16}]$
 & $0.31$($0.70$) & $0.14$($0.25$)
 & $0.15$($0.29$) & $0.18$($0.35$) 
 & $0.19$($0.38$) & $0.14$($0.25$) \\ \hline
$P_{10}^{th}$&$[10^{-20}\,\mbox{GeV}^{-20}]$
 & $0.34$($0.89$) & $0.14$($0.28$) 
 & $0.16$($0.35$) & $0.19$($0.42$) 
 & $0.20$($0.46$) & $0.14$($0.30$) \\ \hline
$P_{20}^{th}$&$[10^{-40}\,\mbox{GeV}^{-40}]$
 & $0.63$($3.75$) & $0.24$($0.80$) 
 & $0.27$($1.12$) & $0.33$($1.36$)  
 & $0.34$($1.54$) & $0.24$($0.92$) \\ \hline\hline
\multicolumn{2}{|c|}{} 
   & \multicolumn{2}{c||}{$\mu_{\rm hard}=5$~GeV}
   & \multicolumn{2}{c||}{$\mu_{\rm soft}=2.5$~GeV}
   & \multicolumn{2}{c|}{$\mu_{\rm soft}=2.5$~GeV} \\ 
\multicolumn{2}{|c|}{} 
   & \multicolumn{2}{c||}{$\mu_{\rm fac}=5$~GeV}
   & \multicolumn{2}{c||}{$\mu_{\rm fac}=5$~GeV}
   & \multicolumn{2}{c|}{$\mu_{\rm hard}=5$~GeV} \\\hline
\end{tabular}
\caption{\label{tabcomp2} 
The theoretical moments $P_n^{th}$ for $n=4,6,8,10,20$ and fixed
$\alpha_s(M_Z)=0.118$ and $M_b^{1S}=4.8$~GeV ($M_b^{pole}=4.8$~GeV)
for various choices of the renormalization scales $\mu_{\rm soft}$,
$\mu_{\rm hard}$ and $\mu_{\rm fac}$. The two-loop running has been
employed for the strong coupling. The values for $P_n^{th}$ in the pole
mass scheme are slightly different from the numbers shown in
Ref.~\cite{Hoang1} because of the correction in  $a_2$.
}
\end{center}
\vskip 3mm
\end{table}
\begin{table}[htb] 
\vskip 7mm
\begin{center}
\begin{tabular}{|c||c|c|c||c|c|c|} \hline
Moment 
   & \multicolumn{3}{|c||}{$M_b^{1S} (M_b^{pole})/[GeV]$}
   & \multicolumn{3}{|c|}{$\alpha_s(M_Z)$} \\ \hline
 & $4.6$ &  $4.8$ & $5.0$ & $0.11$ & $0.12$
                          & $0.13$ \\ \hline\hline 
$P_{4,\mbox{\tiny LO}}^{th}$
 & $0.35$($0.40$) & $0.25$($0.29$) & $0.18$($0.21$)  
 & $0.21$($0.23$) & $0.26$($0.31$) & $0.35$($0.43$)   \\ \hline
$P_{4,\mbox{\tiny NLO}}^{th}$
 & $0.19$($0.26$) & $0.13$($0.18$) & $0.09$($0.13$) 
 & $0.13$($0.16$) & $0.13$($0.18$) & $0.12$($0.23$)  \\ \hline
$P_{4,\mbox{\tiny NNLO}}^{th}$
 & $0.27$($0.39$) & $0.19$($0.28$) & $0.14$($0.20$)
 & $0.17$($0.22$) & $0.20$($0.30$) & $0.26$($0.46$)  \\ \hline
$P_{10,\mbox{\tiny LO}}^{th}$
 & $0.45$($0.62$) & $0.19$($0.27$) & $0.09$($0.12$)
 & $0.15$($0.19$) & $0.21$($0.29$) & $0.30$($0.53$)  \\ \hline
$P_{10,\mbox{\tiny NLO}}^{th}$
 & $0.26$($0.53$) & $0.11$($0.22$) & $0.05$($0.10$) 
 & $0.10$($0.16$) & $0.11$($0.24$) & $0.12$($0.43$) \\ \hline
$P_{10,\mbox{\tiny NNLO}}^{th}$
 & $0.41$($0.90$) & $0.18$($0.39$) & $0.08$($0.17$) 
 & $0.14$($0.25$) & $0.19$($0.44$) & $0.27$($0.94$) \\ \hline\hline 
   & \multicolumn{3}{c||}{\raisebox{-2.5ex}[2.5ex]{$\alpha_s(M_Z) = 0.118$}}
   & \multicolumn{3}{c|}{\raisebox{-.5ex}[.5ex]{$M_b^{1S}=4.8$~GeV}}\\
   & \multicolumn{3}{c||}{}
   &
   \multicolumn{3}{c|}{\raisebox{.5ex}[-.5ex]{($M_b^{pole}=4.8$~GeV)}} \\ \hline
   &      \multicolumn{6}{c|}{$\mu_{\rm soft}=2.5\,\mbox{GeV}\,,\quad
          \mu_{\rm hard}=\mu_{\rm fac}=5$~GeV} \\\hline
\end{tabular}
\caption{\label{tabcomp3} 
The theoretical moments $P_4^{th}$ and $P_{10}^{th}$ at LO, NLO and
NNLO for fixed $\mu_{\rm soft}=2.5$~GeV and 
$\mu_{\rm hard}=\mu_{\rm fac}=5$~GeV for various values of $M_b^{1S}$
($M_b^{pole}$) and $\alpha_s(M_Z)$ in the $1S$ (pole) mass scheme. The
two-loop running  has been employed for the strong coupling. 
The values of $P_{4}^{th}$ and $P_{10}^{th}$ are given in units of
$10^{-8}\,\mbox{GeV}^{-8}$ and $10^{-20}\,\mbox{GeV}^{-20}$, respectively. 
}
\end{center}
\vskip 3mm
\end{table}
\begin{table}[htb] 
\vskip 7mm
\begin{center}
\begin{tabular}{|c||c|c|c|c|} \hline
$n$ & $4$ & $6$ & $8$ & $10$ \\ \hline\hline 
$P_n^{ex}/[10^{-2n}\,\mbox{GeV}^{-2n}]$ & 
    $0.22$ & $0.21$ & $0.23$ & $0.26$ \\ \hline
$(P_n^{ex})^{\Upsilon(1S)}/[10^{-2n}\,\mbox{GeV}^{-2n}]$ & 
    $0.11$ & $0.13$ & $0.17$ & $0.21$ \\ \hline
$(P_n^{ex})^{\Upsilon(1S)}/P_n^{ex}$ & 
    $0.48$ & $0.62$ & $0.72$ & $0.79$ \\ \hline
\end{tabular}
\caption{\label{tabcomp4} 
The experimental moments, $P_n^{ex}$, the contribution of the
$\Upsilon(1S)$ to the experimental moments,
$(P_n^{ex})^{\Upsilon(1S)}$, and the ratio
$(P_n^{ex})^{\Upsilon(1S)}/P_n^{ex}$ for $n=4,6,8,10$.
}
\end{center}
\vskip 3mm
\end{table}
Because the $1S$ mass
is defined purely from the non-relativistic dynamics, the soft scale
has to be employed in Eq.~(\ref{finalfactor}). Thus, the transition to
the $1S$ mass scheme leads to a simple rescaling of the theoretical
moments. We emphasise that the expression in Eq.~(\ref{finalfactor})
must be expanded consistently up to NNLO in the non-relativistic
expansion together with the LO, NLO and NNLO contributions
of $\varrho_{n,1}$ and with $\varrho_{n,2}$  in order to achieve a
proper cancellation of the correlations and the large NNLO corrections
which are present in the pole mass scheme. Like in Ref.~\cite{Hoang1}
we keep the short-distance coefficient $C_1$ in factorized form.

Comparing Eqs.~(\ref{finalfactor}) and (\ref{Pnexpression1}) we can
easily see that the $1S$ scheme represents the most natural scheme one
can use to avoid correlations and large corrections because it reduces
the contribution from the lowest lying bottom-antibottom quark resonance,
which represents the most important part of theoretical moments at
large values of $n$.

Before we turn to the statistical analysis it is quite interesting to
study the impact of the transition to the $1S$ scheme on the
theoretical moments. 
In Tab.~\ref{tabcomp1}
we have displayed the values of $P_n^{th}$ in the $1S$ (pole) scheme
for $n=4,6,8,10,20$ and for various values of $M_b^{1S}$
($M_b^{pole}$) and 
$\alpha_s(M_z)$ while the renormalization scales are fixed to
$\mu_{\rm soft}=2.5$~GeV and $\mu_{\rm hard}=\mu_{\rm fac}=5$~GeV.
The numbers in Tab.~\ref{tabcomp1} show that the dependence of the
moments on the mass is quite strong in the $1S$ as well as in the pole
mass scheme. However, in the $1S$ scheme the variation with respect to
changes in the strong coupling is much weaker, in particular for
larger values of $n$. To illustrate this we have also displayed the
moments for $n=20$, although this value it too high for the practical
application. From this we can expect that the extracted values for
$M_b^{1S}$ are less strongly correlated to the strong coupling
than the pole mass values obtained in Ref.~\cite{Hoang1}. This
feature, however, will also make an independent determination of a
precise value for the strong coupling from the $\Upsilon$ sum rules
impossible in the $1S$ scheme, as we show in the next section.

In Tab.~\ref{tabcomp2} the theoretical moments are displayed for
$n=4,6,8,10,20$ for different choices of $\mu_{\rm soft}$,
$\mu_{\rm hard}$ and $\mu_{\rm fac}$ for $\alpha_s(M_Z)=0.118$
and $M_b^{1S}=4.8$~GeV ($M_b^{pole}=4.8$~GeV) in the $1S$ (pole) mass
scheme.
As expected from the weak dependence of the moments in the $1S$ scheme
on the strong coupling, we also find that the dependence of the
moments on the soft scale is weaker in the $1S$ scheme, in particular
for larger values of $n$. In the $1S$ scheme there is also some
stabilization with respect to variations of the factorisation
scale. The dependence of the moments on the hard scale, on the other
hand, is comparable in both schemes. This is expected because in the
transition from the pole to the $1S$ scheme only non-relativistic
contributions are modified. Because the
variation of the moments with respect to the renormalization scales is
used as an instrument to estimate theoretical uncertainties in the
mass extraction we can expect smaller uncertainties in the $1S$
mass extraction compared to the results shown in Ref.~\cite{Hoang1}.

In Tab.~\ref{tabcomp3} the theoretical moment $P_{4}^{th}$ and
$P_{10}^{th}$ are displayed at LO, NLO and NNLO in the $1S$ (pole) scheme.
Whereas in the pole mass scheme the NNLO corrections are all of
the same size or even larger than the NLO corrections, they are
significantly smaller in the $1S$ scheme. This illustrates the
improvement of the perturbative behaviour of the theoretical
moments if the $1S$ mass scheme is employed. We emphasise, however,
that even in the $1S$ scheme the NNLO
corrections are not {\it per se} small, particularly for large values
of $n$. We believe that this is not a point of concern, because of the
extreme dependence of the moments on the bottom quark mass,
particularly for large values of $n$. It is the behavior of the
results for the mass extracted from the theoretical moments which
should be taken as the measure to judge the quality of the
perturbative expansion.

Finally, in Tab.~\ref{tabcomp4} we have displayed for $n=4,6,8,10$ the
values of the experimental moments $P_n^{ex}$, based on
Eq.~(\ref{Pnexperiment}) and the data given in Tab.~\ref{tabdata}, and
the contribution to $P_n^{ex}$ coming from the $\Upsilon(1S)$ in order
to demonstrate the relative weight of the lowest lying resonance
compared to the other resonances and the continuum. The contribution
of the $\Upsilon(1S)$ is between $48\%$ ($n=4$) and $79\%$
($n=10$). This shows that the higher $\Upsilon$ excitations and the
continuum constitute a significant part of the experimental
moments. Thus, the fact that the value of $M_b^{1S}$ which we determine
in this work (see Eq.~(\ref{M1SNNLO})) is compatible with
$M_{\Upsilon(1S)}/2$ is non-trivial.
\par
\vspace{0.5cm}
\section{Numerical Results} 
\label{sectionnumerics}
To obtain numerical results for the $1S$ mass we use the statistical
procedure described in Ref.~\cite{Hoang1}, which is based on the $\chi^2$
function 
\begin{equation}
\chi^2\Big(M_b^{1S},\alpha_s(M_Z),\mu_{\rm soft},\mu_{\rm
  hard},\mu_{\rm fac},a_{cor}\Big) \, = \,
\sum\limits_{\{n\},\{m\}}\,
\Big(\,P_n^{th}-P_n^{ex}\,\Big)\,(S^{-1})_{n m}\,
\Big(\,P_m^{th}-P_m^{ex}\,\Big)
\,.
\label{x2general}
\end{equation}
$\{n\}$ represents the set of $n$'s for which the fit is carried out
and $S^{-1}$ is the inverse covariance matrix describing the
experimental errors and the correlation between the experimental
moments. The covariance matrix contains the errors in the $\Upsilon$
electronic decay widths, the electromagnetic coupling
$\tilde\alpha_{em}$, and the
continuum cross section $r_{cont}$. The small errors in the $\Upsilon$
masses are neglected. The correlations between the individual
measurements of the electronic decay widths are estimated to be equal
to the product of the respective systematic errors given in
Tab.~\ref{tabdata} times the constant $a_{cor}$. In order to estimate
the theoretical uncertainties in the mass extraction the
renormalization scales $\mu_{\rm soft}$, $\mu_{\rm hard}$, 
$\mu_{\rm fac}$, and the constant $a_{cor}$ are varied randomly in the
ranges
\begin{eqnarray}
1.5\,\mbox{GeV} \, \le & \mu_{\rm soft} & \le \, 3.5\,\mbox{GeV} 
\nonumber\\[1mm]
2.5\,\mbox{GeV} \, \le & \mu_{\rm hard} & \le \, 10\,\mbox{GeV} 
\nonumber\\[1mm]
2.5\,\mbox{GeV} \, \le & \mu_{\rm fac} & \le \, 10\,\mbox{GeV} 
\nonumber\\[1mm]
0 \, \le & a_{cor} & \le \, 1 
\,,
\label{parameterranges}
\end{eqnarray}
and the sets of $n$`s
\begin{equation}
\{n\} \, = \,
\{4,5,6,7\}\,,\{7,8,9,10\}\,,\{4,6,8,10\}
\label{nsets}
\end{equation}
are employed. For each choice of the parameters  $\mu_{\rm soft}$, 
$\mu_{\rm hard}$, $\mu_{\rm fac}$, and $a_{cor}$ and each set of $n$`s
the value of $M_b^{1S}$ is obtained for which $\chi^2$ is minimal. As
in Ref.~\cite{Hoang1} we carry out two types of fits. First, we
consider a fit where $M_b^{1S}$ and $\alpha_s(M_Z)$ are determined
simultaneously (``unconstrained fit''), and, second, we fit for
$M_b^{1S}$ taking $\alpha_s(M_Z)$ as an input (``constrained fit'').

\begin{figure}[t!] 
\begin{center}
\leavevmode
\epsfxsize=4.5cm
\epsffile[220 580 420 710]{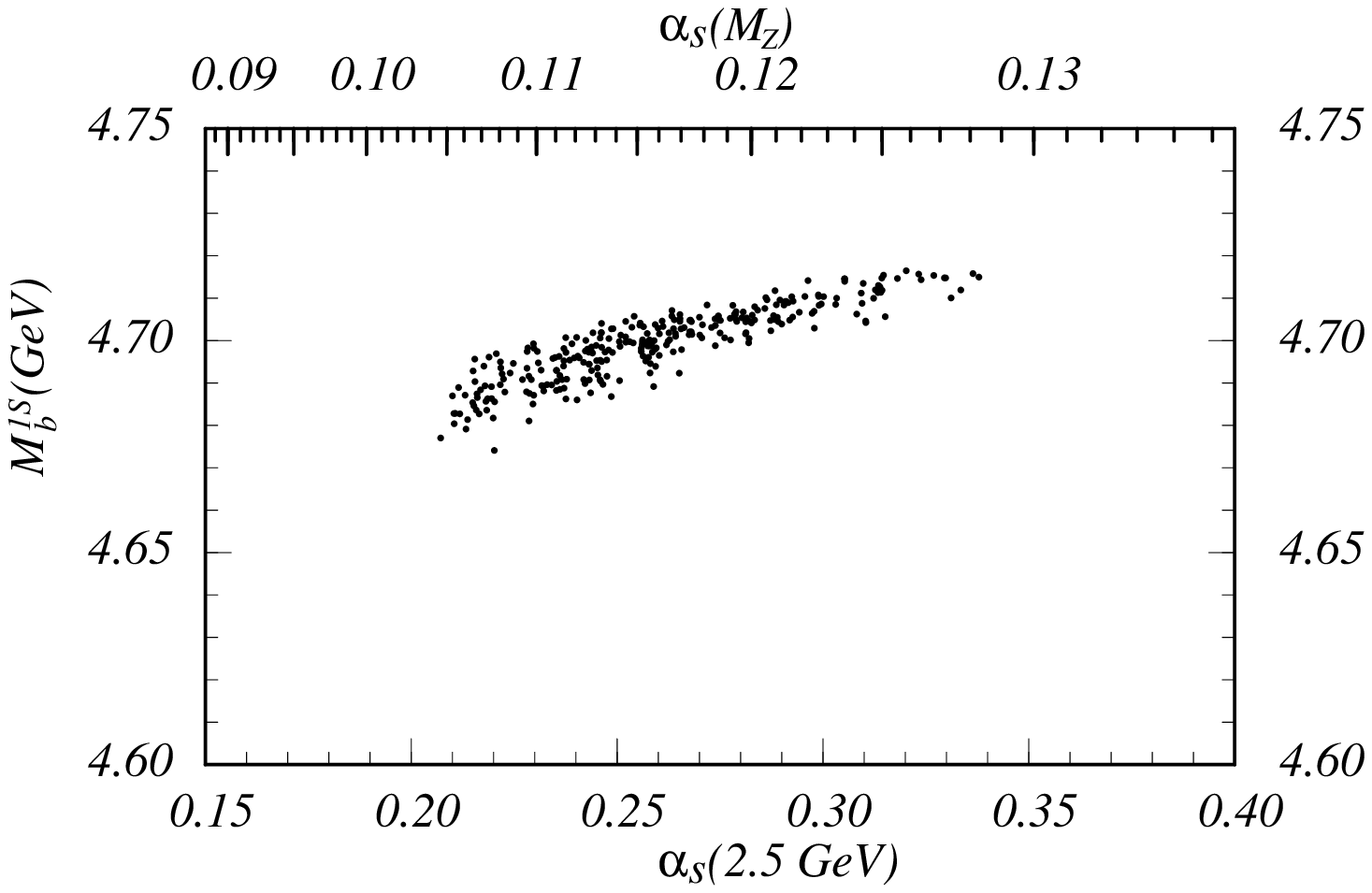}\\
\vskip 3.7cm
\epsfxsize=4.5cm
\leavevmode
\epsffile[220 580 420 710]{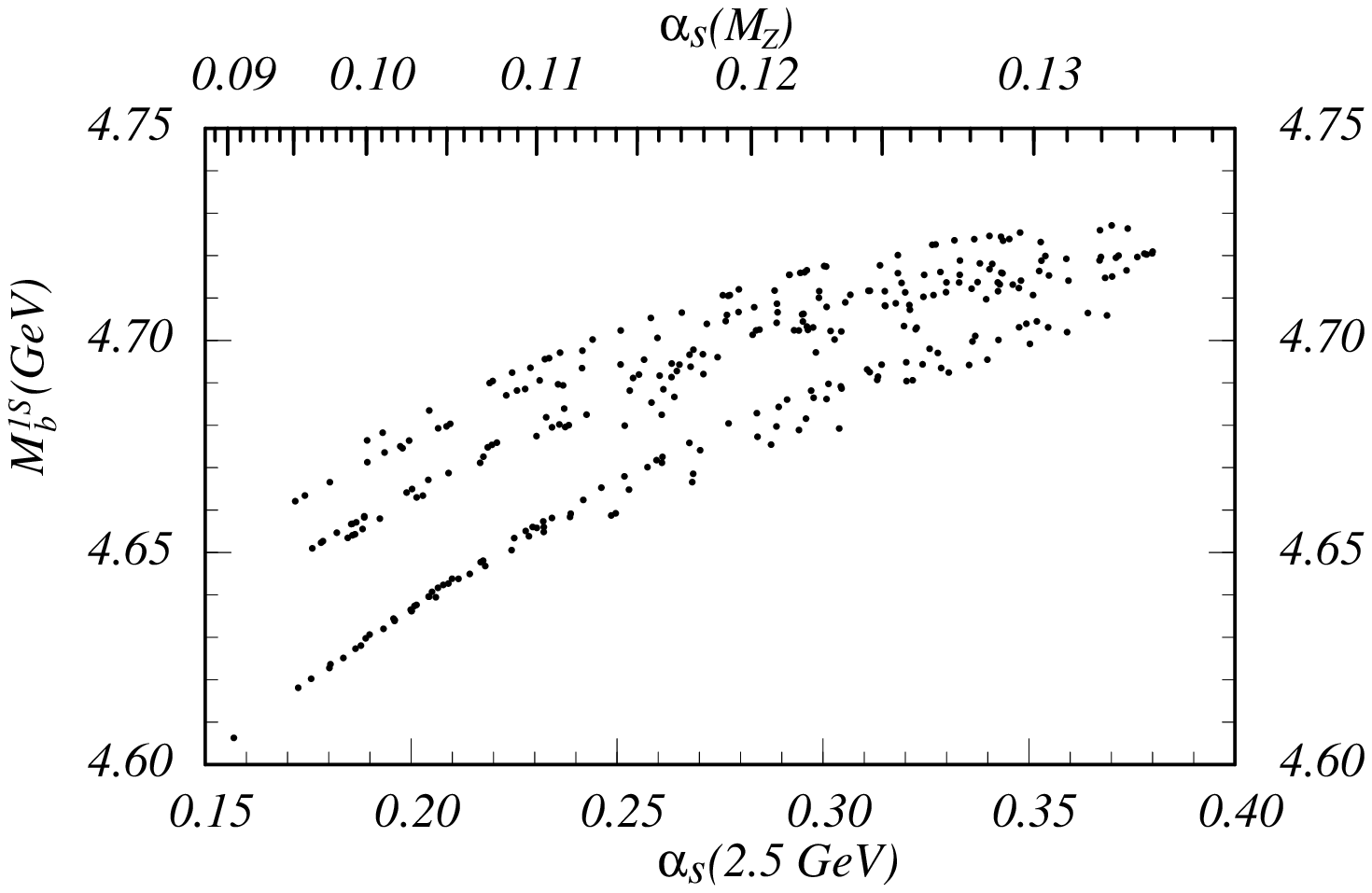}
%
%
\vskip -8.7cm
\mbox{\large ($a$) \hspace{5.5cm}}
\vskip 8.7cm
\vskip -2.55cm
\mbox{\large ($b$) \hspace{5.5cm}}
\vskip 2.55cm
\vskip  2.3cm
 \caption{\label{fignnlouncon} 
Result for the allowed region in the $M_b^{1S}$-$\alpha_s$ plane for the
unconstrained fit based on the theoretical moments at NNLO (a) and NLO
(b). The dots represent points of minimal $\chi^2$ for a large number
of random choices within the ranges~(\ref{parameterranges}) and the
sets~(\ref{nsets}). Experimental errors are not displayed. The two-loop
running has been employed for the strong coupling.
}
 \end{center}
\end{figure}
The result for the allowed region for $M_b^{1S}$ and $\alpha_s(M_Z)$
for the unconstrained fit using the full NNLO expressions for the
theoretical moments is displayed in Fig.~\ref{fignnlouncon}a.
To illustrate the improvement coming from the NNLO corrections
we have also displayed the result for the NLO theoretical
moments in Fig.~\ref{fignnlouncon}b.
The dots represent the points of
minimal $\chi^2$ for a large number of random choices within the
ranges~(\ref{parameterranges}) and the sets~(\ref{nsets}). Comparing
to the corresponding results in the pole mass scheme (see Fig.~10 and
12 in Ref.~\cite{Hoang1}) we find that the range of the mass values
covered by the dots is smaller in the $1S$ scheme. For the NNLO
moments the covered range for the $1S$ mass is $4.67~\mbox{GeV} <
M_b^{1S} < 4.72$~GeV (versus $4.76~\mbox{GeV} <
M_b^{pole} < 4.85$~GeV in the pole scheme~\cite{Hoang1}).
However, the range
of $\alpha_s$ covered in the $1S$ scheme is as large (slightly larger)
as in the pole scheme at NNLO (NLO). This is not unexpected because of
the reduced correlation of the theoretical moments to the strong
coupling mentioned before. 
Obviously, the strong coupling determined from the $\Upsilon$
sum rules in the $1S$ scheme contains uncertainties which are much
larger than the current world averages. We conclude that the
$\Upsilon$ sum rules are not very competitive tool to determine the
strong coupling. We therefore abandon the unconstrained fit also for
the mass determination and turn to the constrained fit.

\begin{figure}[t!] 
\begin{center}
\leavevmode
\epsfxsize=4.5cm
\epsffile[220 580 420 710]{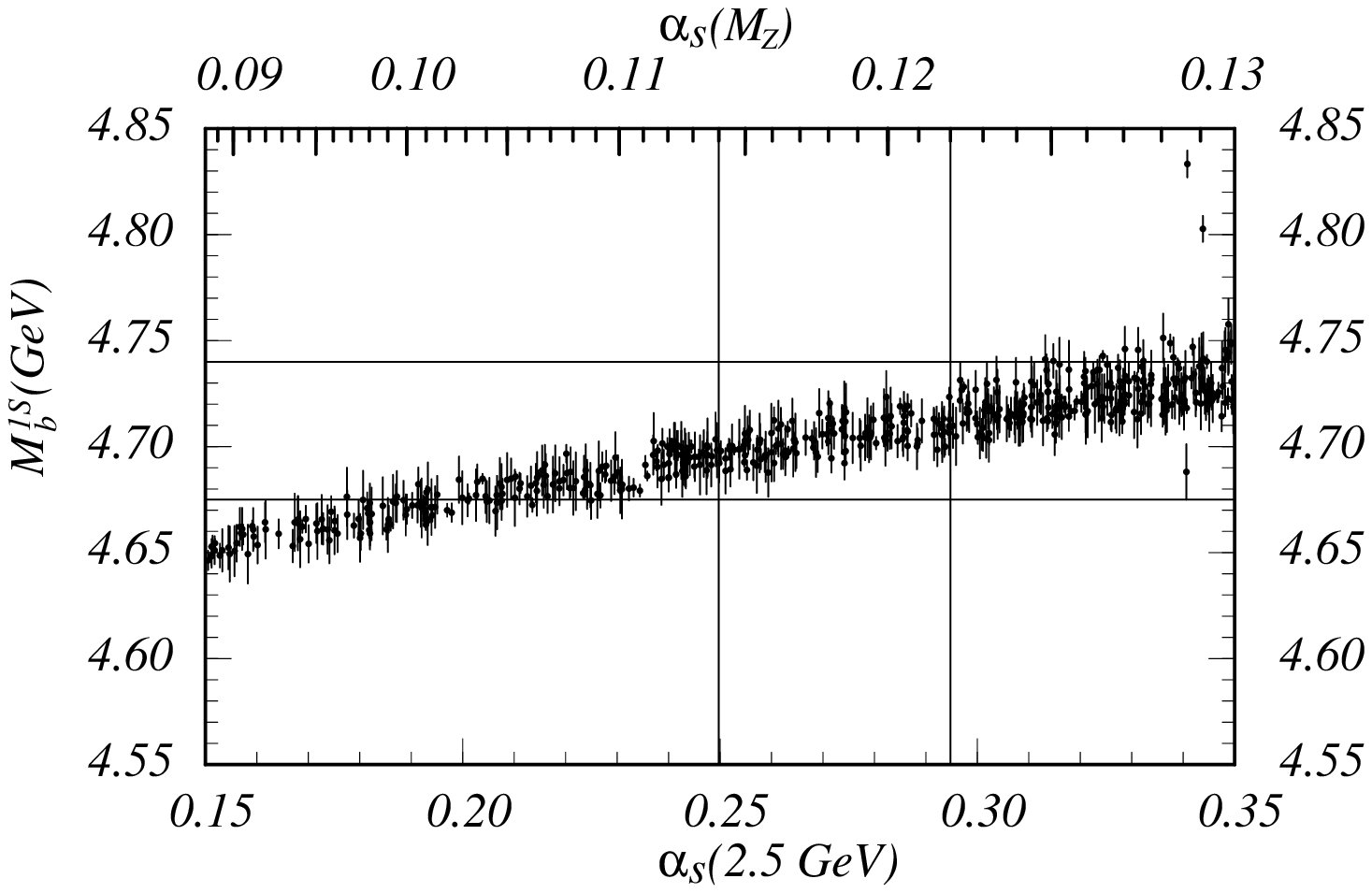}
\vskip 3.7cm
\epsfxsize=4.5cm
\leavevmode
\epsffile[220 580 420 710]{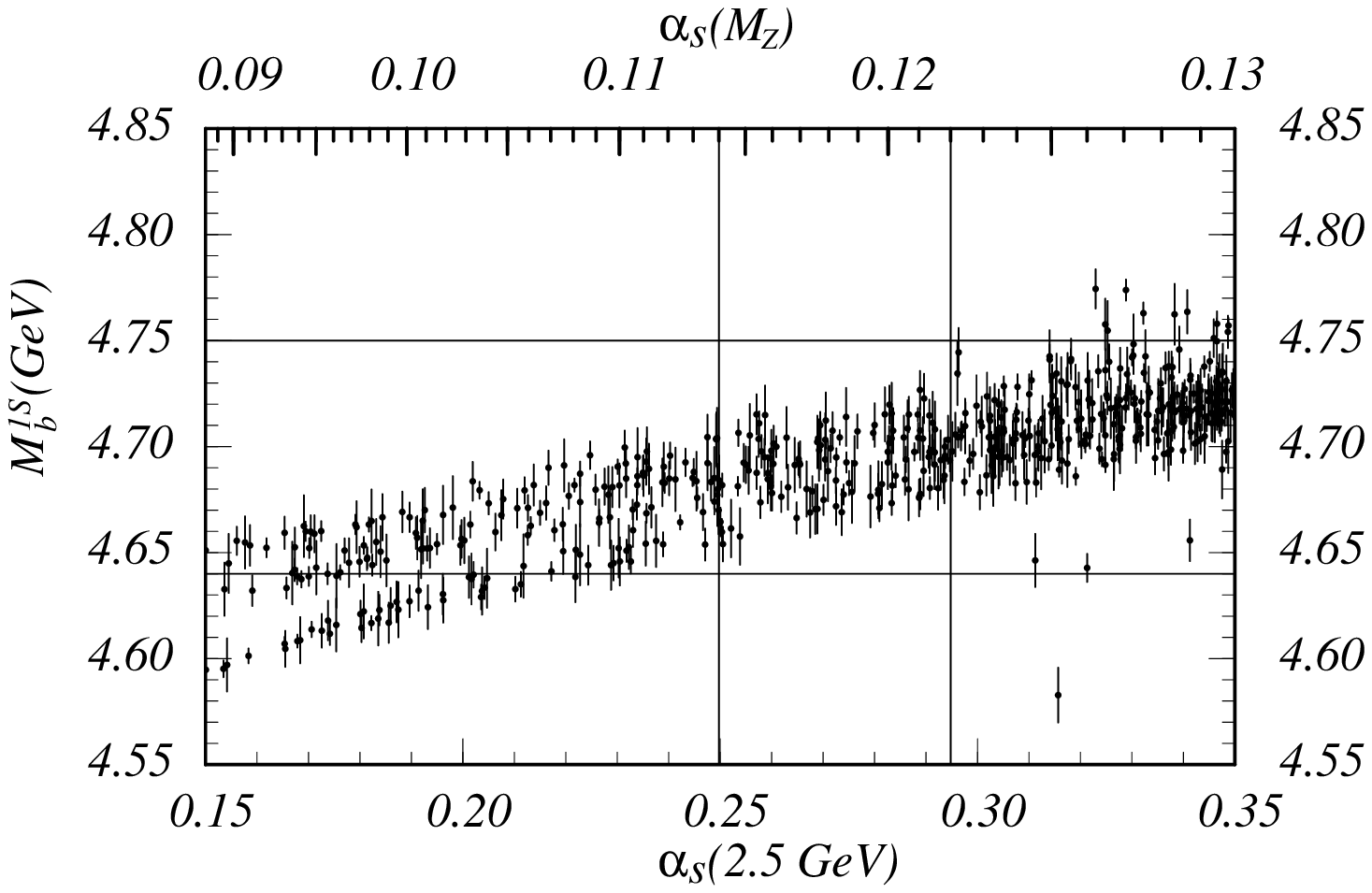}
%
%
\vskip -8.7cm
\mbox{\large ($a$) \hspace{5.5cm}}
\vskip 8.7cm
\vskip -2.55cm
\mbox{\large ($b$) \hspace{5.5cm}}
\vskip 2.55cm
\vskip  2.3cm
 \caption{\label{fignnlocon} 
Result for the allowed $M_b^{1S}$ values for a given value of
$\alpha_s$ at NNLO (a) and NLO (b). The dots represent points of
minimal $\chi^2$ for a large number of random choices within the
ranges~(\ref{parameterranges}) and the sets~(\ref{nsets}), and
randomly chosen values for the strong coupling
$\alpha_s(2.5~\mbox{GeV})$.
Experimental errors at the $95\%$ CL level are displayed as vertical
lines. It is illustrated how the allowed range for $M_b^{1S}$ is
obtained if $0.114\le\alpha_s(M_z)\le 0.122$ is taken as an input. The
two-loop running has been employed for the strong coupling.
}
 \end{center}
\end{figure}
In Fig.~\ref{fignnlocon}a the allowed range for $M_b^{1S}$ is displayed
as a function of $\alpha_s$ based on the NNLO theoretical
moments. Fig.~\ref{fignnlocon}b shows the result of the same analysis
using the NLO theoretical moments only. Each dot represents the $1S$ mass
for which the $\chi^2$ function is minimal for a given input value of
$\alpha_s$, and for a random choice within the
ranges~(\ref{parameterranges}) and the sets~(\ref{nsets}).
The statistical errors
corresponding to $95\%$~CL are below $15$~MeV for all the dots
displayed in Figs.~\ref{fignnlocon}a and b.
In contrast to the constrained fit results in the pole scheme 
(see Fig.~13 in Ref.~\cite{Hoang1}) we find that in the
$1S$ scheme the correlation between the mass and the strong coupling
is reduced considerably. Compared to the pole mass results the
uncertainties are smaller by a factor three for the NLO as well as the
NNLO analyses. Comparing Fig.~\ref{fignnlocon}a
and b we also see that the NNLO contributions to the moments reduce
the uncertainties of the NLO analysis roughly by a factor two.
Taking $\alpha_s(M_z)=0.118\pm 0.004$ as input we arrive at 
\begin{equation}
M_b^{1S} \, = \, 4.71 \pm 0.03~\mbox{GeV}
\label{M1SNNLO}
\end{equation} 
for the $1S$ mass obtained from the NNLO theoretical moments, where
also the experimental errors have been taken into account. This result
is consistent with the unconstrained fit shown in
Fig.~\ref{fignnlouncon}a. 
The NLO analysis yields $M_b^{1S} \, = \, 4.70 \pm 0.06$~GeV.
Due to the small correlation of the $1S$ mass to the strong coupling
the result does not change significantly if a slightly different
choice of the 
central value and the error of $\alpha_s(M_Z)$ is assumed.
Eq.~(\ref{M1SNNLO}) represents the main result of this paper. We note
that the choice of the range of the renormalization scale  
$\mu_{\rm soft}$ has the main impact on the uncertainty in the $1S$
mass determination. We have checked that allowing for larger values of 
$\mu_{\rm soft}$ does not affect the results displayed in
Figs.~\ref{fignnlocon}. However, the uncertainties are larger if
values of $\mu_{\rm soft}$ smaller than $1.5$~GeV are admitted. If a
lower bound for $\mu_{\rm soft}$ of $1$~GeV is chosen, the result for
the $1S$ mass reads $M_b^{1S}=4.73\pm 0.05$~GeV using the NNLO
theoretical moments. However, we believe that $\mu_{\rm soft}\ge
1.5$~GeV is already a conservative choice.

To conclude this section we would like to note that our method to
extract the $1S$ bottom quark mass based on the $\chi^2$ function for
several moments (see Eq.(\ref{x2general})) leads to smaller
uncertainties than if the fit would be carried out for individual
moments independently. This can be easily seen if the scale dependence
of the moments shown in Tab.~\ref{tabcomp2} is compared to their mass
dependence shown in Tab.~\ref{tabcomp1}. The $\chi^2$ function has a
smaller scale dependence than the individual moments because it puts
higher weight on linear combinations of the moments which are more
sensitive to the relative size than to the absolute size of the
moments. These linear combinations are determined from the entries of
the covariance matrix (see Eqs.~(81) and (82) of Ref.~\cite{Hoang1})
which account for the correlation of the experimental moments in
Eq.~(\ref{Pnexperiment}) coming from their dependence on the
$\Upsilon$ masses and electronic widths.  
\par
\vspace{0.5cm}
\section{Determination of the $\overline{\mbox{MS}}$ Mass
and the Upsilon Expansion}
\label{sectionmsbar}
The $1S$ mass is, by construction, optimally adapted to reduce the
dependence of the theoretical moments on the strong coupling and the
renormalization scale $\mu_{\rm soft}$, which govern the
non-relativistic dynamics of the bottom-antibottom quark pair. By the
same token, the $1S$ mass does not know much about large momenta
above the inverse Bohr radius $\sim M_b\,\alpha_s$. Thus, although the
$1S$ mass can, like the $\overline{\mbox{MS}}$ mass, serve as a proper
short-distance mass
definition in its own right, it can only serve as a specialised
short-distance mass definition for systems where the characteristic
momentum scale is comparable to the inverse Bohr radius.\footnote{
By considering the fictitious case of inclusive semileptonic $B$
decays with the radiation of $d$ additional massless colour-singlet
scalars at the weak vertex, where $d$ is large, it was pointed out in
Ref.~\cite{Bigi2} that the characteristic scale for the inclusive $B$
decay rates is $M_b/(5+2d)$ rather than $M_b$. Because
$M_b/(5+2d)\approx M_b\,\alpha_s$ for $d=0$, this might be an
explanation for the drastic improvement of the convergence behaviour
in the low order perturbative series for the inclusive $B$ decay rates
in the $1S$ scheme which has been reported in Ref.~\cite{Hoang3,Hoang4}.
}
For high energy processes, where momenta of order $M_b$ and higher are
the characteristic scales, the $\overline{\mbox{MS}}$ mass is much
better adapted. Thus, it is mandatory to determine the
$\overline{\mbox{MS}}$ mass from the value of the $1S$ mass given in
Eq.~(\ref{M1SNNLO}). We emphasise again that using specialised
short-distance masses can reduce perturbative uncertainties because
correlations are minimised. This is a consequence of the fact that
usually only a few low orders of perturbation theory are known. 

Great care has to be taken if the $\overline{\mbox{MS}}$ mass is
expressed in terms of the $1S$ mass in order to ensure the proper
cancellation of the most IR sensitive terms which exist in the
perturbative series which expresses both masses in terms of
$M_b^{pole}$. It has been shown in Ref.~\cite{Hoang3,Hoang4} that the
consistent
way to relate the $1S$ and the $\overline{\mbox{MS}}$ mass to each
other is the upsilon expansion. The upsilon expansion states that
the LO, NLO and NNLO non-relativistic contributions in
Eq.~(\ref{M1Sdef}) are of order $\epsilon$, $\epsilon^2$ and
$\epsilon^3$, respectively, of the auxiliary parameter $\epsilon=1$
mentioned already in Sec.~\ref{section1Smass}. On the other hand, in
the relation which expresses the pole mass in terms of the
$\overline{\mbox{MS}}$ mass the terms of order $\epsilon$,
$\epsilon^2$ and $\epsilon^3$ in the upsilon expansion correspond to
the one-, two- and three-loop contributions, respectively. When the 
$\overline{\mbox{MS}}$ mass is expressed in terms of the $1S$ mass we
then have to use the expansion in $\epsilon$ instead of
$\alpha_s$. This means that at each order in $\epsilon$ different
orders of $\alpha_s$ are mixed together. This unusual prescription can
be understood from the fact that the most IR sensitive contributions
in Eq.~(\ref{M1Sdef}) (i.e. those contributions which involve the
highest power of $\beta_0$ in each order) involve powers of the
logarithmic term $L=\ln(\mu/C_F\,a_s\,M_b^{pole})$. These logarithmic
terms exponentiate at larger orders, 
$\sum_{i=0}L^i/i!\approx \exp(L)=\mu/C_F\,a_s\,M_b^{pole}$, and
effectively cancel one power of $\alpha_s$~\cite{Hoang3,Hoang4}. We
will show below that
this exponentiation is already very effective at NNLO in
Eq.~(\ref{M1Sdef}) making the use the upsilon expansion mandatory to
achieve a reliable determination of the $\overline{\mbox{MS}}$ mass.

The upsilon expansion states that we need the three-loop contribution
in the relation between the $\overline{\mbox{MS}}$ and the pole mass
in order to be able to make use of the known NNLO contribution
$\Delta^{NNLO}$ in Eq.~(\ref{M1Sdef}). For the purpose of illustration
we will use in the following the large-$\beta_0$ approximation for the
yet unknown three-loop contributions. The relation between the pole
and the $\overline{\mbox{MS}}$ mass then reads~\cite{Broadhurst1,Ball1}
\begin{eqnarray}
M_b^{pole} & = & \overline m_b(\overline m_b)\,\bigg[\,
1 + 0.424\,\alpha_s(\overline m_b)\,\epsilon + 
    0.158\,\alpha_s^2(\overline m_b)\,(\beta_0-2.384)\,\epsilon^2 +
    0.047\,\alpha_s^3(\overline m_b)\beta_0^2\,\epsilon^3
\,\bigg]
\,.
\label{poleMSbar}
\end{eqnarray}
Combining Eq.~(\ref{M1Sdef}) with Eq.~(\ref{poleMSbar}) using the
upsilon expansion up to $\epsilon^3$ we arrive at the following result
for the $\overline{\mbox{MS}}$ bottom quark mass
\begin{eqnarray}
 \overline m_b(\overline m_b) & = &
\bigg[\,
4.71\,-\,0.40\,\epsilon\,-\,0.11\,\epsilon^2\,-\,
0.04\,\epsilon^3\,\pm\,0.03\,\Big(\delta M_b^{1S}\Big)
\,\pm\,x\,0.01\,\Big(\delta\alpha_s\Big)
\,\bigg]\,\mbox{GeV}
\label{MSbarestimate}
\end{eqnarray}
where we have taken Eq.~(\ref{M1SNNLO}) as input for the $1S$ mass and
assumed $\alpha_s(M_z)=0.118\pm x\,0.001$ for the strong coupling. If
the actual three-loop contribution in Eq.~(\ref{poleMSbar}) is just
$10\%$ smaller than the large-$\beta_0$ approximation, the
$\epsilon^3$ term in Eq.~(\ref{MSbarestimate}) is smaller by a factor
of two. This illustrates the effectiveness of the exponentiation of
the logarithmic terms mentioned above already at NNLO in
Eq.~(\ref{M1Sdef}). We emphasise that, because of this large
cancellation, the $\overline{\mbox{MS}}$ mass would have a large
(positive) systematic error if the NNLO contribution in
Eq.~(\ref{M1Sdef}) would be used without also including the three-loop
contributions in Eq.~(\ref{poleMSbar}). This systematic shift would
amount to about $+200$~MeV.\footnote{
This systematic error is included in results given in
Ref.~\cite{Pineda1}.
}
Thus at present stage we have to leave out \underline{all}
$\epsilon^3$ terms 
completely, and, assuming an uncertainty in $\alpha_s(M_Z)$ of $0.004$
($x=4$), we arrive at
\begin{eqnarray}
\overline m_b(\overline m_b) & = & 4.20 \, \pm \, 0.06~\mbox{GeV} 
\label{MSbarfinal}
\end{eqnarray} 
for the $\overline{\mbox{MS}}$ bottom quark mass. To obtain the error
in Eq.~(\ref{MSbarfinal}) we have added all errors quadratically
assuming a perturbative uncertainty due to ignorance of the exact size
of the $\epsilon^3$ terms of $40$~MeV. We believe that our optimistic
assumption about the perturbative uncertainty is justified in view of
the good quality of the large-$\beta_0$ approximation which can be
expected at the three-loop level in Eq.~(\ref{poleMSbar}).
In order to reduce the uncertainty in the $\overline{\mbox{MS}}$
bottom quark mass the three-loop contributions in the relation between
the $\overline{\mbox{MS}}$ and the pole mass must be determined
\underline{and} the precision of the strong coupling has to be
improved.
\par
\vspace{0.5cm}
\section{The Low Scale Running and the Potential Subtracted Bottom
  Quark Masses}
\label{sectionother}

In recent literature there have been two other proposals for
specialised short-distance mass definitions which can be used for 
a stable and correlation-reduced bottom quark mass determination from the
$\Upsilon$ sum rules. 

In Refs.~\cite{Voloshin2,Bigi2} the ``low scale
running mass'', $M_b^{LS}$, was proposed to subtract the low momentum
behaviour of the bottom quark self energy in the pole mass
scheme.\footnote{
In this language the $1S$ mass just subtracts half the perturbative
binding energy of a ${}^3\!S_1$ bottom-antibottom quark bound state in
the pole mass scheme. 
} 
Like the $1S$ mass the
``low scale running mass'' was originally devised to improve
the convergence of the perturbative contributions in inclusive $B$
meson decays. Apart from the
renormalization scale which governs the strong coupling the low scale
running mass depends, in addition, on the cutoff $\mu_{LS}$ which limits
the momenta subtracted from the self energy. If this cutoff is
adjusted to be close to the inverse Bohr radius the low scale running
mass acts very similar as the $1S$ mass. The relation between 
$M_b^{LS}$ and the pole mass is known to two-loop
order~\cite{Czarnecki1}, which
corresponds to order $\epsilon^2$ in the upsilon
expansion. In order to allow for compatibility checks for sum rule
analyses and other bottom quark mass determinations within the framework
of the low scale running mass scheme, it is useful to determine $M_b^{LS}$
from the $1S$ mass obtained in this work.
Taking the large-$\beta_0$ approximation for the yet
unknown three-loop (order $\epsilon^3$ in the upsilon
expansion) contribution, following Ref.~\cite{Melnikov1}, and using
again the upsilon expansion one can determine $M_b^{LS}(\mu_{LS})$
from our result for the $1S$ mass in Eq.~(\ref{M1SNNLO}). For
$\mu_{LS}=1$~GeV the result reads
\begin{eqnarray}
M_b^{LS}(\mu_{LS}=1~\mbox{GeV}) & = & \bigg[\,
4.71\,-\,0.09\,\epsilon\,-\,0.05\,\epsilon^2\,-\,
0.02\,\epsilon^3\,
\nonumber\\[2mm]
& & \mbox{\hspace{1cm}}
\pm\,0.03\,\Big(\delta M_b^{1S}\Big)
\,\pm\,x\,0.003\,\Big(\delta\alpha_s\Big)
\,\pm\,0.01\,\Big(\mu\Big)
\,\bigg]~\mbox{GeV}
\,,
\label{MLSestimate}
\end{eqnarray}
assuming again $\alpha_s(M_Z)=0.118\pm x\,0.001$ for the strong
coupling as input and choosing $\mu=5$~GeV as the
renormalization scale of the strong coupling. The third uncertainty in
Eq.~(\ref{MLSestimate}) arises from the variation of 
this renormalization scale of the strong coupling between $2.5$ and
$10$~GeV. In the relation between the $1S$ and the low scale running
mass it is reasonable to choose the renormalization scale of order the
bottom quark mass because the physical effects involving the inverse
Bohr radius are eliminated. [Allowing for $\mu=1.5$~GeV would not change
the result shown in Eq.~(\ref{MLSestimate}).] 
The uncertainty from $\alpha_s$ in Eq.~(\ref{MLSestimate}) (and also in
Eq.~(\ref{MPSestimate})) is smaller
than in the corresponding uncertainty in Eq.~(\ref{MSbarestimate})
because the order $\epsilon$ term is much smaller in
Eq.~(\ref{MLSestimate}) (and in Eq.~(\ref{MPSestimate})). 
The relation between $M_b^{LS}$ and $M_b^{1S}$ is very well
behaved illustrating the fact the both are short-distance masses. The
size of the scale uncertainty is compatible with the size of the
$\epsilon^3$ term. Assuming an uncertainty of $0.004$ ($x=4$) for 
$\alpha_s(M_Z)$ and taking into account that the exact $\epsilon^3$
contribution is not yet known we arrive at 
\begin{eqnarray}
M_b^{LS}(\mu_{LS}=1~\mbox{GeV}) & = & 4.57\pm 0.04~\mbox{GeV}
\label{MLSfinal}
\end{eqnarray}
for the low scale running mass at $1$~GeV. As in the previous section
we have combined the error coming from $M_b^{1S}$, $\alpha_s$,
and a perturbative uncertainty of $20$~MeV quadratically. 
The result for the low scale running mass in Eq.~(\ref{MLSfinal}) is
compatible with a recent determination of
$M_b^{LS}(\mu_{LS}=1~\mbox{GeV})$ which was also based on the
$\Upsilon$ sum rules at NNLO ($M_b^{LS}=4.56\pm
0.06$~GeV)~\cite{Melnikov1}. The analysis of Ref.~\cite{Melnikov1},
however, is different from ours with respect
to several aspects. In Ref.~\cite{Melnikov1} the energy denominators
in the cross section $R(s)$ have been resummed for energies close to
the resonances, and the final dispersion integration~(\ref{momentsdef2})
has been carried out numerically, which means that the moments
calculated in Ref.~\cite{Melnikov1} contain certain non-relativistic
contributions beyond NNLO. In this work the energy denominators are
not resummed and all integrations are carried out analytically,
consistently dropping all higher order non-relativistic contributions
beyond NNLO (see Ref.~\cite{Hoang1} for details). This means that our
analysis only uses global duality arguments, whereas
Ref.~\cite{Melnikov1} also relies on the validity of local duality. In
addition, in  Ref.~\cite{Melnikov1} larger values of $n$ ($n=14-18$)
and larger values of the soft scale $\mu_{\rm soft}$ 
($\mu_{\rm soft}=2-4.5$~GeV) where used, and $M_b^{LS}$ was determined
from extracting values for the pole mass first and converting them
into numbers for the low scale running mass afterwards. In our
analysis the pole mass has been eliminated completely. We also point
out that in Ref.~\cite{Melnikov1} the values of $M_b^{LS}$ were
obtained by fitting individual moments whereas
in this work a $\chi^2$ function based on several moments was
employed (see the comment at the end of
Sec.~\ref{sectionnumerics}). The compatibility of Eq.~(\ref{MLSfinal})
with the result obtained in Ref.~\cite{Melnikov1} might serve as an
argument that the methodical differences in this analysis and the one
in Ref.~\cite{Melnikov1} do not affect the mass determination.

In Ref.~\cite{Beneke2} the ``potential
subtracted'',$M_b^{PS}$, mass was proposed. It subtracts the low
momentum contribution of the static potential in the pole mass
scheme~\cite{Fischler1,Billoire1,Schroeder1}. Like the low
scale running mass the potential subtracted mass depends on a cutoff,
$\mu_{PS}$. For $\mu_{PS}=\frac{4}{3}\mu_{LS}$ the low scale running
and the potential subtracted mass are approximately equal. This
equivalence is based on the universality of the most infrared
sensitive contributions contained in the corresponding
subtractions. So far there has not been any 
sum rule analysis which has attempted to extract the potential
subtracted bottom quark mass. However, it is useful to determine
$M_b^{PS}$ from the $1S$ mass value obtained in this work to allow for
cross checks with possible future work on this subject. Because the
potential 
subtracted mass is based on the static perturbative potential, which
is known to NNLO in the non-relativistic expansion (i.e. $\epsilon^3$ in
the upsilon expansion), the perturbative uncertainties in the relation
between potential subtracted and $1S$ mass are smaller.
Starting from our result for the $1S$ mass, Eq.~(\ref{M1SNNLO}),
the potential subtracted mass at the subtraction scale
$\mu_{PS}=2$~GeV reads  
\begin{eqnarray}
M_b^{PS}(\mu_{PS}=2~\mbox{GeV}) & = & \bigg[\,
4.71\,-\,0.13\,\epsilon\,-\,0.04\,\epsilon^2\,-\,
0.01\,\epsilon^3\,
\nonumber\\[2mm]
& & \mbox{\hspace{1cm}}
\pm\,0.03\,\Big(\delta M_b^{1S}\Big)
\,\pm\,x\,0.003\,\Big(\delta\alpha_s\Big)
\,\pm\,0.01\,\Big(\mu\Big)
\,\bigg]~\mbox{GeV}
\,,
\label{MPSestimate}
\end{eqnarray}
where we have again assumed $\alpha_s(M_Z)=0.118\pm x\,0.001$ as an
input for the strong coupling. The third uncertainty arises from
the variation of the renormalization scale in the strong coupling
between $2.5$ and $10$~GeV. The relation between $M_b^{PS}$ and
$M_b^{1S}$ is very well behaved illustrating the fact that $M_b^{PS}$
is also a short-distance mass. Assuming again an uncertainty of
$0.004$ ($x=4$) for $\alpha_s(M_Z)$, taking a perturbative uncertainty
of $10$~MeV and adding all errors quadratically we arrive at 
\begin{eqnarray}
M_b^{PS}(\mu_{PS}=2~\mbox{GeV}) & = & 4.53\pm 0.03~\mbox{GeV}
\label{MPSfinal}
\end{eqnarray}
for the potential subtracted  mass at $2$~GeV. 

To conclude this section we note that it is possible to tune the
cutoff scales in the low scale running and the potential subtracted
masses, at each order of perturbation theory, such that the
correlations of the mass to other parameters are minimal. Such a
fine-tuning approach, however, is illegal because from the conceptual
point of view any cutoff scale close to the inverse Bohr radius is
equally well suited. This ``scale-ambiguity'' is a specific
characteristic of the low scale running and the potential subtracted
mass and not inherent in the $1S$ mass, because the latter is defined
through a bound state mass, which is a physical quantity. We therefore
consider the $1S$ scheme as the most natural mass scheme to be used to
describe processes involving heavy quark-antiquark pairs in the
non-relativistic regime. We emphasise, however, that the $LS$, $PS$ and
$1S$ mass improve the quality of the mass extraction from the
$\Upsilon$ sum rules all by the same mechanism, and that the most
infrared sensitive contributions contained in the subtractions are
equivalent in all schemes. In the $1S$ scheme the
cutoff which is visible in the definition of the $LS$ and the $PS$
masses is provided in a natural way by the width of the
wave-function of the ${}^3\!S_1$ bottom-antibottom quark bound state.
No matter which scheme is used the final answers for the
$\overline{\mbox{MS}}$ bottom quark mass should be compatible.
\par
\vspace{0.5cm}
\section{Conclusions}
\label{sectionconclusions}
In this work we have extracted the $1S$ mass, $M_b^{1S}$, from sum
rules which relate the masses and electronic decay widths of the
$\Upsilon$ mesons to large-$n$ moments of the vacuum polarization
function, which have been calculated at NNLO in the non-relativistic
expansion. $M_b^{1S}$ is defined as half the perturbative mass of a
$J^{PC} = 1^{--}$, ${}^3\!S_1$ bottom-antibottom quark bound state,
i.e. it represents the perturbative contribution of half the mass of
the $\Upsilon(1S)$ meson. However, the latter information is not used in
this work and $M_b^{1S}$ is considered as a fictitious mass
parameter which is determined solely from the sum rule analysis.
The result for the $1S$ mass reads $M_b^{1S}=4.71\pm 0.03$~GeV, if
$\alpha_s(M_Z)=0.118\pm 0.004$ is taken as an input for the strong
coupling, and has only a small correlation to the choice of $\alpha_s$.
The close proximity of this result to
$M_{\Upsilon(1S)}/2$ is non-trivial because higher excitations and the
$b\bar b$ continuum have a non-negligible contribution to the sum rules
for the values of $n$ used in this work. The fact that, within the
errors, $M_b^{1S}$ is equal to $M_{\Upsilon(1S)}/2=4.730$~GeV
indicates that
the non-perturbative contributions in the ${\Upsilon(1S)}$ mass are
probably small. Compared to an earlier work which has been carried out
in the pole mass scheme by the same author we find an improved
perturbative behaviour of the moments and a much smaller residual
dependence on the renormalization scale which governs the
non-relativistic dynamics of the bottom-antibottom quark pair.
The $1S$ bottom quark mass is a short-distance mass (i.e. it does not
have an ambiguity of order $\Lambda_{QCD}$) and can be used as
a mass parameter in its own right. It is well adapted to processes
where the characteristic scale is below the bottom quark mass and has
been successfully applied earlier to inclusive $B$ mesons decays,
where it leads to a significant improvement in the behaviour of the
perturbative series describing the inclusive
widths.~\cite{Hoang3,Hoang4}  

Because $M_b^{1S}$ is a short-distance mass it has a well behaved
perturbative relation to the $\overline{\mbox{MS}}$ mass, which is
adapted to processes where the characteristic scale is of order the
bottom quark mass or higher. In order to relate the $1S$ mass to the 
$\overline{\mbox{MS}}$ mass definition it is mandatory
to use the upsilon expansion~\cite{Hoang3,Hoang4} in order to ensure
the cancellation of large infrared-sensitive contributions present in
the perturbative series which relate both masses to the pole
mass. Using the presently known two-loop contributions in the relation
of the $\overline{\mbox{MS}}$ and the pole mass (i.e. using the
upsilon expansion up to order $\epsilon^2$) and assuming
$\alpha_s(M_Z)=0.118\pm 0.004$ we arrive at
$\overline m_b(\overline m_b)=4.20\pm 0.06$~GeV for the
$\overline{\mbox{MS}}$ mass. The error in the $\overline m_b(\overline
m_b)$ can be reduced once the three-loop contributions in
the relation between the $\overline{\mbox{MS}}$ and the pole mass are
calculated and if the precision in $\alpha_s$ is increased. 
\par
\vspace{.5cm}
\section*{Acknowledgements}
I thank Z. Ligeti, A. Manohar and M. B. Voloshin for comments to
the manuscript, and M. Beneke and K. Melnikov for useful conversation.
This work is supported in part by the EU Forth Framework Program
``Training and Mobility of Researchers'', Network ``Quantum
Chromodynamics and Deep Structure of Elementary Particles'', contract
FMRX-CT98-0194 (DG12-MIHT). 

\vspace{1.0cm}
%
\sloppy
\raggedright
\def\app#1#2#3{{\it Act. Phys. Pol. }{\bf B #1} (#2) #3}
\def\apa#1#2#3{{\it Act. Phys. Austr.}{\bf #1} (#2) #3}
\def\lhc{Proc. LHC Workshop, CERN 90-10}
\def\npb#1#2#3{{\it Nucl. Phys. }{\bf B #1} (#2) #3}
\def\nP#1#2#3{{\it Nucl. Phys. }{\bf #1} (#2) #3}
\def\plb#1#2#3{{\it Phys. Lett. }{\bf B #1} (#2) #3}
\def\prd#1#2#3{{\it Phys. Rev. }{\bf D #1} (#2) #3}
\def\pra#1#2#3{{\it Phys. Rev. }{\bf A #1} (#2) #3}
\def\pR#1#2#3{{\it Phys. Rev. }{\bf #1} (#2) #3}
\def\prl#1#2#3{{\it Phys. Rev. Lett. }{\bf #1} (#2) #3}
\def\prc#1#2#3{{\it Phys. Reports }{\bf #1} (#2) #3}
\def\cpc#1#2#3{{\it Comp. Phys. Commun. }{\bf #1} (#2) #3}
\def\nim#1#2#3{{\it Nucl. Inst. Meth. }{\bf #1} (#2) #3}
\def\pr#1#2#3{{\it Phys. Reports }{\bf #1} (#2) #3}
\def\sovnp#1#2#3{{\it Sov. J. Nucl. Phys. }{\bf #1} (#2) #3}
\def\sovpJ#1#2#3{{\it Sov. Phys. LETP }{\bf #1} (#2) #3}
\def\jl#1#2#3{{\it JETP Lett. }{\bf #1} (#2) #3}
\def\jet#1#2#3{{\it JETP Lett. }{\bf #1} (#2) #3}
\def\zpc#1#2#3{{\it Z. Phys. }{\bf C #1} (#2) #3}
\def\ptp#1#2#3{{\it Prog.~Theor.~Phys.~}{\bf #1} (#2) #3}
\def\nca#1#2#3{{\it Nuovo~Cim.~}{\bf #1A} (#2) #3}
\def\ap#1#2#3{{\it Ann. Phys. }{\bf #1} (#2) #3}
\def\hpa#1#2#3{{\it Helv. Phys. Acta }{\bf #1} (#2) #3}
\def\ijmpA#1#2#3{{\it Int. J. Mod. Phys. }{\bf A #1} (#2) #3}
\def\ZETF#1#2#3{{\it Zh. Eksp. Teor. Fiz. }{\bf #1} (#2) #3}
\def\jmp#1#2#3{{\it J. Math. Phys. }{\bf #1} (#2) #3}
\def\yf#1#2#3{{\it Yad. Fiz. }{\bf #1} (#2) #3}
\def\ufn#1#2#3{{\it Usp. Fiz. Nauk }{\bf #1} (#2) #3}
\def\spu#1#2#3{{\it Sov. Phys. Usp.}{\bf #1} (#2) #3}
\def\epjc#1#2#3{{\it Eur. Phys. J. C }{\bf #1} (#2) #3}

\end{document}